\documentclass[twocolumn,showpacs,preprintnumbers,amsmath,amssymb]{revtex4-1}

\usepackage{graphicx}
\usepackage{dcolumn}
\usepackage{bm}
\usepackage{hyperref}
\usepackage{amsmath}

\begin{document}


\title{Chimera death induced by the mean-field diffusive coupling} 



\author{Tanmoy Banerjee}
\email{Electronic mail: tbanerjee@phys.buruniv.ac.in}
\affiliation{Department of Physics, University of Burdwan, Burdwan 713 104, West Bengal, India.}


\received{:to be included by reviewer}
\date{\today}

\begin{abstract}
Recently a novel dynamical state, called the {\it chimera death}, is discovered in a network of non locally coupled identical oscillators [A. Zakharova, M. Kapeller, and E. Sch\"{o}ll, Phy.Rev.Lett. 112, 154101 (2014)], which is defined as the coexistence of spatially  coherent and incoherent oscillation death state. This state arises due to the interplay of non locality and symmetry breaking and thus bridges the gap between two important dynamical states, namely the chimera and oscillation death. In this paper we show that the chimera death can be induced in a network of generic identical oscillators with mean-field diffusive coupling and thus we establish that a non local coupling is not essential to obtain chimera death. We identify a new transition route to the chimera death state, namely the transition from in-phase synchronized oscillation to chimera death via global amplitude death state. We ascribe the occurrence of chimera death to the bifurcation structure of the network in the limiting condition and show that multi-cluster chimera death states can be achieved by a proper choice of initial conditions.
\end{abstract}


\maketitle 

Cooperative phenomena in a network of coupled oscillators have been an active topic of extensive research in the field of physics, biology, engineering and social science \cite{sync}.  Two fascinating cooperative dynamical states, namely the {\it chimera} and {\it oscillation death} have been in the center of recent research and a burst of publications have explored many aspects of their origin and manifestations \cite{kosprep} \cite{chireview}. 

The chimera is an intriguing spatio-temporal dynamical state  where the synchronous and asynchronous behavior are observed simultaneously in a network of coupled identical oscillators \cite{kuro}. After its discovery in phase oscillators \cite{kuro} the chimera state attracts immediate attention due to its surprisingly complex behavior and a possible connection with many real world phenomena such as unihemispheric sleep in certain species \cite{chireview}, the multiple time scales of sleep dynamics \cite{neurochi}, etc. In the initial years the chimera state was found in the phase part under non local coupling, but recently it has been observed in many other coupling configurations, also \cite{sethia1,*sethia2} \cite{nonglobal,*danachi} (see \cite{chireview} and references therein). Further, in the strong coupling limit it is found that amplitude effects come into play that result in amplitude mediated chimera \cite{sethia1,*sethia2} and amplitude chimera states \cite{scholl2,*scholl3}. The existence of chimera in real systems has been established  experimentally in optical \cite{raj}, chemical \cite{show} and electronic systems \cite{fortuna}.

On the other hand, oscillation death (OD) is an oscillation quenching state where coupling dependet symmetry breaking of a network gives rise to stable inhomogeneous steady states \cite{kosprep}.  \citet{kosprl} established that OD is a different state in comparison with its homogeneous counterpart, namely the amplitude death state (AD), where all the oscillators arrive at a common homogeneous steady state. OD has a strong connection and importance in the field of biology (e.g., synthetic genetic oscillator \cite{kosepl,*koschaos}, cellular differentiation \cite{cell}) and physics \cite{odphys}. The recent works reported the occurrence of OD and several transition routes from AD to OD \cite{kosprl} under different coupling schemes \cite{kurthpre}, e.g.,  dynamic and conjugate coupling \cite{kospre}, time-delay coupling \cite{scholl4}, repulsive coupling \cite{dana,*dana1}, mean-field coupling \cite{tanpre1,tanpre2} and direct-indirect coupling \cite{tanarxiv}. 

Although both {\it chimera} and {\it oscillation death} state induce inhomogeneity in a rather homogeneous network of oscillators, but their inter connection was not identified until the  recent pioneering work by \citet{scholl}. In \cite{scholl} the authors reported an important discovery of a new state, which they called the {\it chimera death}, and established the much speculated connection between the chimera and oscillation death. According to Ref.\cite{scholl},  {\it chimera death} (CD) is the steady state version of chimera, i.e., the population of oscillators in a network splits into incongruous coexisting domains of spatially coherent OD (where neighboring nodes attain essentially the same branch of the inhomogeneous steady state) and spatially incoherent OD (where the neighboring nodes jump among the different branches of inhomogeneous steady state in a completely random manner). They argued that chimera death arises in non locally coupled Stuart-Landau oscillators due the {\it interplay of non locality and symmetry breaking}. They further show a transition from amplitude chimera  to CD state via in-phase synchronized oscillatory state and an abrupt direct transition in the strong coupling limit.

In this paper we address the following open questions: is it possible to induce {\it chimera death} in a network of identical oscillators with a coupling topology other than non local coupling? If yes, what underlying principle is responsible for that? We indeed identify that the mean-field diffusive coupling is able to induce chimera death in a network of identical oscillators and thus establish that non locality is not an essential ingredient to induce chimera death. The mean-field coupling is widely studied in the field of biology, physics, and engineering and previously shown to induce AD and OD in identical oscillators \cite{tanpre1,tanpre2,qstr,*srimali,*tanchaosad}. For our study we choose the paradigmatic Stuart-Landau oscillator and van der Pol oscillator separately and identify two possible transitional routes to chimera death: (i) a transition from  in-phase synchronized oscillation to chimera death via  global amplitude death state; this transition scenario is reported for the first time and has not been observed in Ref.\cite{scholl}; (ii) a direct abrupt transition from in-phase synchronized oscillation to chimera death. We show that the occurrence of either of this transitions depends upon the coupling strength and the density of mean-field and has a connection with the bifurcation structure of the network in the limiting case of two oscillators. Finally, we show that multi-cluster chimera death can be induced by the proper choice of initial conditions.

We consider a network of $N$ Stuart-Landau oscillators interacting through mean-field diffusive coupling; mathematical model of the coupled system is given by
\begin{equation}\label{ls} 
\dot{Z_i}=(1+j\omega_i-|Z_i|^{2})Z_i+\epsilon\bigg(Q\overline{Z}-Re(Z_i)\bigg).
\end{equation}
Here $i=1\cdots N$; $\overline{Z}=\frac{1}{N}\sum_{i=1}^{N}Re(Z_i)$ is the mean-field of the coupled system, $Z_i=x_i+jy_i$. Each oscillators have a unit amplitude and an eigenfrequency $\omega_i$. The coupling strength is given by $\epsilon$. $Q$ is the density of mean-field which controls the influence of the mean-field on the system dynamics; it is a relevent parameter in many systems in the field of biology (e.g., genetic oscillators interacting through a quorum-sensing mechanism) and physics \cite{qstr,*srimali,*tanchaosad}, $ 0\leqslant Q \leqslant 1$. 

\begin{figure}
\includegraphics[width=.45\textwidth]{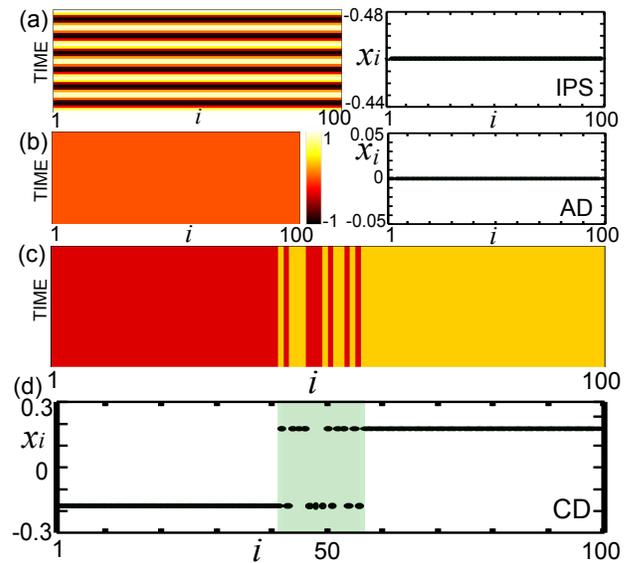}
\caption{\label{fig1}(Color online) Spatio-temporal color map and snapshot of amplitude plots for (a) in-phase synchronized oscillation (IPS) at $\epsilon=2$ (b) global amplitude death (AD) at $\epsilon=4$ (c-d) chimera death (CD) at $\epsilon=8$. The spatially incoherent OD region is shown with filled color for visual guidance in (d). A transient of $t=5000$ is excluded. Other parameters are: $Q=0.4$, $\omega=2$.}
\end{figure}

We examine the spatio-temporal dynamics of a network of $N=100$ mean-field coupled Stuart-Landau oscillators with the variation of coupling strength $\epsilon$ and density of mean-field $Q$ (without any loss of generality we take $\omega=2$ \cite{scholl}). Numerical simulations are carried out with the $\mbox{XPPAUT}$ package \cite{xpp}; we use the fourth-order Runge-Kutta method (step size$=0.01$) and left out a large number of initial iterations ($t=5000$) to exclude the transient behavior. Fig.~\ref{fig1} shows the spatial pattern of the network for different coupling strength ($\epsilon$) at $Q=0.4$. We observe that at lower (non zero) value of coupling strength all the oscillators in the network are in-phase synchronized with each other. This is shown in Fig.~\ref{fig1} (a) with $\epsilon=2$. With increasing $\epsilon$, beyond a certain value we observe a {\it global} amplitude death state, i.e., now all the oscillators arrive at the common trivial steady state (which is zero in the present case) and oscillations cease to take place. This is shown in Fig.~\ref{fig1} (b) for $\epsilon=4$. Further increase in coupling strength results in {\it chimera death}, which is shown in Fig.~\ref{fig1} (c-d) for $\epsilon=8$. In this state the network consists of two distinct coexisting regions: in one region we have {\it spatially coherent} branches of OD state with the inhomogeneous steady states $x^*\approx\pm0.176$ ($x^*$ is derived later in this paper) and in the other region we observe {\it spatially incoherent} OD state, i.e., in this region oscillators populate either of the two branches of OD in a random sequence. This is equivalent to the {\it chimera death} state observed in Ref.\cite{scholl} in non locally coupled Stuart-Landau oscillators. From Fig.~\ref{fig1} (c-d) we can see that the first 40 oscillators (i.e., $x_{1-40}$) attain a value $x^{*}_{1-40}\approx-0.176$, oscillators indexed $56-100$ have value $x^{*}_{56-100}\approx0.176$, but oscillators indexed $41-55$ populates either $x^\ast\approx0.176$ or $x^\ast\approx-0.176$ in a random sequence [see the filled region of Fig.~\ref{fig1}(d)]. Therefore, with increasing coupling strength ($\epsilon$), we identify the transition from in-phase synchronized oscillation (IPS) [Fig.~\ref{fig1} (a)] to global amplitude death [Fig.~\ref{fig1} (b)] to chimera death state [Fig.~\ref{fig1} (c-d)]. This transition route to chimera death was not observed earlier in Ref.\cite{scholl}.

\begin{figure}
\includegraphics[width=.45\textwidth]{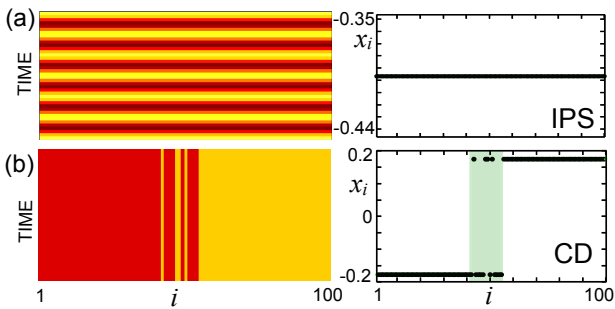}
\caption{\label{fig2} (Color online) Spatio-temporal color map (color code same as Fig.~\ref{fig1}) (left panel) and snap shot of amplitude plots (right panel) for (a) in-phase synchronized oscillation (IPS) at $\epsilon=4$ (b) chimera death (CD) at $\epsilon=8$. A transient of $t=5000$ is excluded. Other parameters are: $Q=0.7$, $\omega=2$.}
\end{figure}

We identify one more transition route, namely the direct abrupt transition from in-phase synchronized oscillation (IPS) to chimera death that occurs for a higher value of $Q$. Figure~\ref{fig2} shows this for the variation of $\epsilon$ with $Q=0.7$. Fig.~\ref{fig2} (a) depicts the in-phase synchronized oscillation for $\epsilon=4$ and Fig.~\ref{fig2} (b) shows the chimera death state for a higher value, $\epsilon=8$. In Ref.\cite{scholl} a direct abrupt transition from amplitude chimera (AC) to CD state was observed. Unlike Ref.\cite{scholl}, we find no AC state in our system (however, for lower coupling strength, we observe AC in the transient part that takes a long time before it decays to an IPS state), {\it thus we may infer that the occurrence of amplitude chimera is not a prerequisite to achieve chimera death}.

\begin{figure}
\includegraphics[width=.45\textwidth]{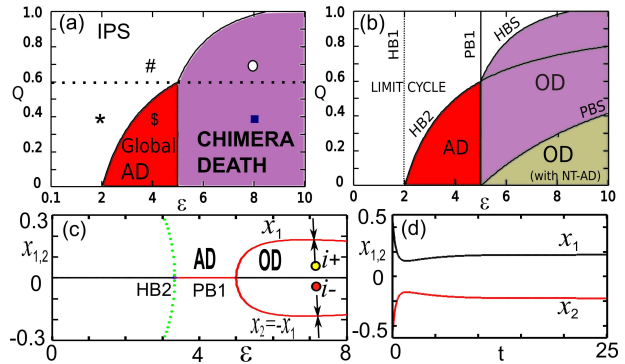}
\caption{\label{fig3} (Color online) (a) Phase diagram in $\epsilon-Q$ space ($\omega=2$). The $\ast$, {\$} and $\blacksquare$ represent the parameter values for generating Fig.\ref{fig1} (a), (b) and (c-d), respectively; $\#$ and $\circ$ represent that for Fig.\ref{fig2} (a) and (b), respectively. (b) Two parameter bifurcation diagram of two mean-field coupled Stuart-Landau oscillators. NT-AD is a non trivial AD state that coexists with OD in certain zone of parameter space. (c) One dimensional bifurcation diagram. Gray (red) [black] line indictes stable [unstable] steady state. (d) Real time evolution of $x_1(t)$ and $x_2(t)$ ($\epsilon=8$): initial conditions are $x_1(0)=0.5$ and $x_2(0)=-0.5$ (also, $y_{1,2}(0)=\mp0.5$). In (c-d) $Q=0.4$, $\omega=2$.}
\end{figure}

Next we map the dynamics of the network in $\epsilon-Q$ parameter space. Figure~\ref{fig3}(a) shows this for $\omega=2$. We find that there exists a critical value $Q^*$ ($=0.6$) that separates two types of transition scenarios. For $Q<Q^*$ the first type of transition (i.e., transition from IPS to chimera death via global AD) occurs and for $Q>Q^*$ the latter transition (from IPS to chimera death) occurs. This is suggestive enough to examine the following question: are these routes to chimera death related with the bifurcation structure of the mean-field coupled Stuart-Landau oscillators? We indeed find the answer in affirmative. The phase diagram of Fig.~\ref{fig3}(a) has a striking resemblance to the two-parameter bifurcation diagram of two mean-field coupled Stuart-Landau oscillators that was discussed in detail in Ref.\cite{tanpre1} and is shown in Fig.~\ref{fig3}(b). In particular,  the OD region of bifurcation diagram [Fig.~\ref{fig3}(b)] matches exactly with the chimera death region of Fig.~\ref{fig3}(a) that indicates the direct relation between this two states. To better understand the scenario let us consider the limiting case of $N=2$ oscillators. From Eq. \eqref{ls}, apart from the trivial $(0,0,0,0)$ fixed point, one has an inhomogeneous steady state (${x_1}^\ast$, ${y_1}^\ast$, $-{x_1}^\ast$, $-{y_1}^\ast$) stabilization of which gives OD; where ${x_1}^\ast = -\frac{\omega {y_1}^\ast}{{\omega}^2 + \epsilon {{y_1}^\ast}^2}$ and ${y_1}^\ast = \sqrt {\frac{(\epsilon - 2{\omega}^2) + \sqrt{{\epsilon}^2 - 4{\omega}^2}}{2\epsilon}}$. From Fig.~\ref{fig3}(b), it can be seen that for $Q<Q^*$, HB2 curve organizes the transition from oscillation state to AD via supercritical Hopf bifurcation and the PB1 line determines the transition from AD to OD through pitchfork bifurcation. The locus of HB1 and PB1 was derived in \cite{tanpre1} as:
\begin{subequations}
\label{bif}
\begin{align}
\label{hb2}
{\epsilon}_{HB2} &= \frac{2}{1-Q},\\
\label{pb1}
{\epsilon}_{PB1} &= 1+{\omega}^2.
\end{align}
\end{subequations}
PB1 and HB2 collide at $ Q^\ast = \frac{{\omega}^2-1}{{\omega}^2+1}$ ($=0.6$ for $\omega=2$) and  for $Q>Q^*$, a subcritical Hopf bifurcation is responsible for the direct {\it abrupt} transition from oscillation state to the OD state (and thus from IPS to chimera death state), which is determined by the HBS curve [Fig.~\ref{fig3}(b)] given by \cite{tanpre1}
\begin{equation}
\label{hbs}
{\epsilon}_{HBS} = \frac{-2(Q+1)+4\sqrt{1+{\omega}^2(1-Q)(3+Q)}}{(1-Q)(3+Q)}.
\end{equation}
A direct {\it abrupt} transition from amplitude chimera to the chimera death state was observed in Ref.~\cite{scholl}, but its origin was not explored there. 

\begin{figure}
\includegraphics[width=.4\textwidth]{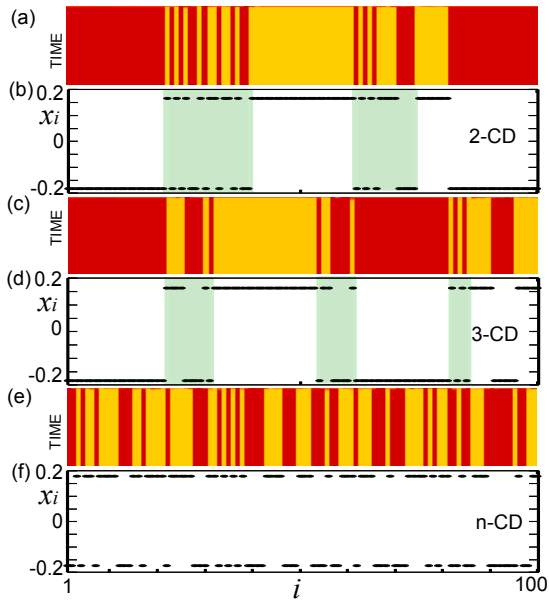}
\caption{\label{fig4} (Color online) Initial condition dependent multi-cluster chimera death states for $\epsilon=8$ (color code same as Fig.~\ref{fig1}). (a-b) two-cluster chimera death (2-CD) (c-d) three-cluster chimera death (3-CD) (e-f) $n$-cluster chimera death ($n$-CD). Gray (color) filled regions in (b) and (d) are for the visual guidance of the incoherent OD regions. Other parameters are: $Q=0.4$, $\omega=2$.}
\end{figure}

In Ref.~\cite{scholl} a multi-cluster chimera death state was observed depending upon the ratio of coupling range to the number of oscillators ($N$). In our present case we find that, for a particular set of initial conditions, the spatial pattern of chimera death state remains the same for all $\epsilon$ and $Q$ values in the ``chimera death" region of Fig.~\ref{fig3}(a). Also, we verify that the spatial position of the incoherent OD region does not vary with $N$, although the organization of the coherent OD regions depends upon $N$. Thus, multi-cluster chimera death state can not be induced in this system by simply changing the parameter values. However, we observed that several multi-cluster chimera death states can be achieved by using judiciously prepared initial conditions. Figure.~\ref{fig4}(a-b) show a two-cluster chimera death state, while Fig.~\ref{fig4}(c,d) and (e,f) show a three-cluster chimera death and an $n$-cluster chimera death, respectively; in all the three cases coupling strength is same ($\epsilon=8$) but initial conditions are different (with $Q=0.4$ and $\omega=2$). For example, in  Figure.~\ref{fig4} (a-b), a choice of linearly increasing and decreasing spatial initial conditions results in the coherent OD states, but for the two incoherent OD regions we choose  random initial conditions that are distributed uniformly in the range $(-0.5,0.5)$ (with mean=0). In the case of three-cluster CD [Fig.~\ref{fig4} (c-d)], for the three incoherent OD regions we choose  random spatial initial conditions with the same distribution. For the last case ($n$-cluster CD) we choose random spatial initial conditions for all the oscillators. Thus, we observe that, in the OD region, a proper choice of spatial initial conditions can induce several multi-cluster CD states. 

This dependence upon initial conditions can be intuitively understood from the one dimensional bifurcation diagram of two Stuart-Landau oscillators that is shown in Figure.~\ref{fig3} (c). Here, beyond the pitchfork bifurcation (PB1) one has two symmetric branches of OD, the upper branch is $x_1$ (say) and thus the lower branch is $x_2=-x_1$. A positive initial condition for $x_1$ (say $i+$) and a negative initial condition for $x_2$ (say $i-$)  (and also a proper choice of the initial conditions of  $y_{1,2}$) populate the upper branch with $x_1$ [see the real time trace of $x_{1,2}$ in Fig.~\ref{fig3}(d) as an example]. Thus, in the OD region if we choose random initial conditions (symmetrically around zero) for both $x_1$ and $x_2$ then they populate the two branches in a random manner (with the following constraint: $x_2=-x_1$). Intuitively, this simple argument may be extended to a large number of oscillators, also: e.g., in a network of oscillators, if we choose a set of random initial conditions (symmetrically around zero) for a certain set of oscillators, then it is expected that set of oscillators will populate the upper and lower branch in a random manner and form a spatially incoherent OD state. However, this argument is true only qualitatively because the exact bifurcation scenario of a large number of oscillators is very much complex and also the final state depends upon the dynamics of the system, thus the final dynamical state resulted from a set of initial conditions is difficult (if not impossible) to predict. Nevertheless, this simple argument has worked in our case and is instructive to explore the role of initial conditions in inducing chimera death in other coupling configurations, also. 

\begin{figure}
\includegraphics[width=.48\textwidth]{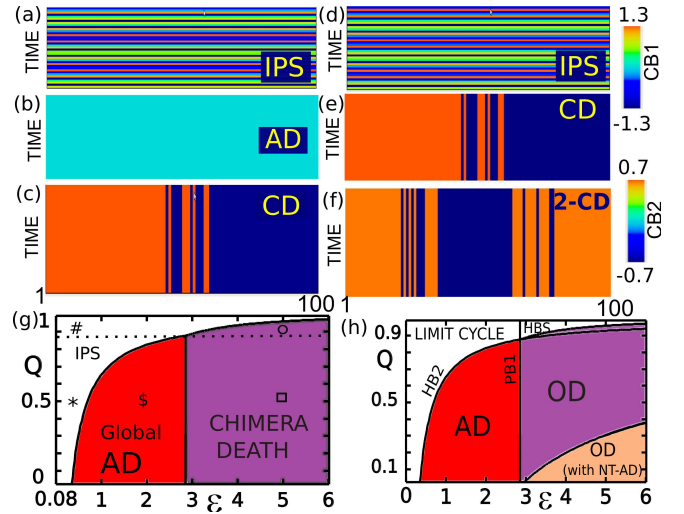}
\caption{\label{fig5} (Color online) Results for the network of van der Pol oscillators ($a=0.35$) using $x$ variable. (a-c) Transition from IPS ($\epsilon=0.3$) to global AD ($\epsilon=2$) to chimera death ($\epsilon=5$) for $Q=0.5$ ($<Q^*$). (d,e) Direct transition from IPS ($\epsilon=0.3$) to CD ($\epsilon=5$) for $Q=0.9$ ($>Q^*$). (f) Initial condition dependent multi-cluster chimera death state (a 2-CD state) for $\epsilon=5$, $Q=0.5$. Colorbar CB1 is for (a,b,d) and CB2 is for (c,e,f). (g) Phase diagram in $\epsilon-Q$ space. The $\ast$, {\$} and $\Box$ represent the parameter values used for generating figures of (a), (b) and (c), respectively; $\#$ and $\circ$ represent that for (d) and (e), respectively. (b) Two parameter bifurcation diagram of two mean-field coupled van der Pol oscillators.}
\end{figure}

Next, to verify the generality of all the results obtained for the network of Stuart-Landau oscillators we repeat this study for a network of another paradigmatic limit cycle oscillator, namely the van der Pol oscillator. We consider $N$ van der Pol oscillators interacting through mean-field diffusive coupling described by
\begin{subequations}
\label{system}
\begin{align}
\label{x1}
\dot{x}_{i} &= y_i+\epsilon\left(Q\overline{X}-x_{i}\right),\\
\label{y1}
\dot{y}_{i} &= a(1-x_i{^2})y_i-x_i.
\end{align}
\end{subequations}
$i=1\cdots N$; $\overline{X}=\frac{1}{N}\sum_{i=1}^{N}x_i$ is the mean-field of the coupled system. The individual oscillators show a near sinusoidal oscillation for a small $a$ ($>0$). Through an extensive numerical study (with $N=100$ and $a=0.35$) we find that all of the following phenomena that we have observed for the Stuart-Landau network are preserved for the network of van der Pol oscillators, also: e.g., the occurrence of chimera death, two transition routes to CD depending upon $Q$, the correspondence between the phase diagram and two-parameter bifurcation diagram, and initial condition dependent multi-cluster CD state. We present the results in Fig.\ref{fig5}. For $Q=0.5$, Fig.\ref{fig5}(a-c) show the transition from in-phase synchronized oscillation (IPS) ($\epsilon=0.3$) to global AD ($\epsilon=2$) to chimera death ($\epsilon=5$) with increasing coupling strength. However, for $Q=0.9$ we observe  the direct transition from IPS ($\epsilon=0.3$) to CD ($\epsilon=5$) [Fig.\ref{fig5}(d,e)]. We also induce several multi-cluster chimera death states using properly prepared spatial initial conditions; as an example, Fig.\ref{fig5}(f) shows a two-cluster CD (2-CD) state (for $\epsilon=5$ and $Q=0.5$). Like the Stuart-Landau network, here also we identify the correspondence between the phase diagram [Fig.\ref{fig5}(g)] of the network and the two parameter bifurcation diagram of two van der Pol oscillators [Fig.\ref{fig5}(h)]. The critical $Q$ value ($Q^*$) that separates two different transition scenarios can be obtained from the bifurcation point, where HB2 and PB1 collide. The locus of PB1 and HB2 were derived in \cite{tanpre2} and are given by: ${\epsilon}_{PB1} = \frac{1}{a}$ and ${\epsilon}_{HB2} = \frac{a}{1-Q}$, respectively . Thus, we get $ Q^\ast = (1- a^2)$ ($=0.8775$ for $a=0.35$) and this matches exactly with the phase diagram of the network [Fig.\ref{fig5}(g)].

In conclusion, we have shown that the chimera death state can be induced in a network of mean-field coupled generic oscillators and thus established that the  non local coupling is not an essential requirement for inducing this state. We have identified a new transition route to the chimera death state, namely the transition from in-phase oscillation to chimera death via global amplitude death state, which was not observed in the non local coupling case \cite{scholl}; also we have explored the role of mean-field density parameter in governing the transitions. We have established the correspondence between the chimera death state in a network and the two-parameter bifurcation diagram of the network in the limiting case of two oscillators. We further qualitatively explained the role of initial conditions in inducing multi-cluster chimera death state. We believe that this study will initiate the search for the chimera death in other coupling schemes that are not non local in topology and at the same time open up the research to engineer multi-cluster chimera death state in electronic and biological networks.

The author acknowledges the assistance of Ms. Debarati Ghosh in the initial phase of this work. This work is supported by SERB, Department of Science and Technology (DST), India [project grant: SB/FTP/PS-005/2013].   



\providecommand{\noopsort}[1]{}\providecommand{\singleletter}[1]{#1}%

\end{document}